\documentclass[structabstract]{aa}

\usepackage{psfrag}

\usepackage{graphicx}
\usepackage{txfonts}
\begin{document}
   \title{The intrinsic dispersion of the Faber-Jackson relation for early-type galaxies as function of the mass and redshift}
   \titlerunning{The intrinsic dispersion of the FJR as function of the mass and redshift}

   \authorrunning{Nigoche-Netro et al.}



   \author{A. Nigoche-Netro  \inst{1}, J. A. L. Aguerri \inst{2,3}, P. Lagos \inst{2,3}, A. Ruelas-Mayorga \inst{4}, L. J. S\'anchez \inst{4}, C. Mu\~noz-Tu\~n\'on \inst{2,3} \and  A. Machado \inst{4} }


   \institute{Instituto de Astrof\'isica de Andaluc\'ia (IAA), Glorieta de la Astronom\'ia s/n, 18008, Granada, Spain. \\ \email{nigoche@iaa.es} \and Instituto de Astrof\'isica de Canarias (IAC), V\'ia L\'actea s/n, 38200 La Laguna, Spain. \\ \email{jalfonso;plagos@iac.es}
\and Departamento de Astrof\'isica, Universidad de La Laguna, C/ Molinos de Agua s/n, 38205, La Laguna, Spain.
\and Instituto de Astronom\'ia, Universidad Nacional Aut\'onoma de M\'exico, Apartado Postal 70-264,  Cd Universitaria 04510,  M\'exico D.F., M\'exico.\\  \email{rarm;leonardo;machado@astroscu.unam.mx} }


  \abstract
   {}
   {Recently it has been reported that the intrinsic dispersion at constant magnitude of the structural relations from early-type galaxies is a useful tool to study the universality of these structural relations, that is to say, to study whether the structural relations depend on luminosity, wavelength, redshift and/or environment. In this work we study the intrinsic dispersion of the Faber-Jackson relation as function of the luminosity, mass and redshift.}
   {We use a sample of approximately 90 000 early-type galaxies from the Sloan Digital Sky Survey (SDSS-DR7) spanning a magnitude range of 7 $mag$ in both $g$ and $r$ filters. We calculate the intrinsic dispersion of the Faber-Jackson relation at approximately constant magnitude and compare this at different luminosities, masses and redshifts.}
   {The main results are the following:  i) The intrinsic dispersion of the Faber-Jackson relation depends on the luminosity, mass and redshift. ii) The distribution for brighter and more massive galaxies has smaller intrinsic dispersion than that for fainter and less massive galaxies. iii) The distribution of bright and massive galaxies at higher redshift has smaller intrinsic dispersion than those similar galaxies at low redshift. }
   {Comparisons of the results found in this work with recent studies from the literature make us conclude that the intrinsic dispersion of the Faber-Jackson relation could depend on the history of galaxies, in other words, the intrinsic dispersion could depend on the number and nature of transformation events that have affected the galaxies along their life times, such as collapse, accretion, interaction and merging. }

   \keywords{Galaxies:fundamental parameters, photometry, distances and redshifts}

   \maketitle
%

\section{Introduction}

The scale relations of early-type galaxies (ETGs) are mathematical relations involving the galaxies' structural parameters such as: effective radius ($r_{e}$), mean effective surface brightness inside $r_{e}$ ($\left\langle \mu\right\rangle _{e}$), central velocity dispersion ($\sigma_0$) and total absolute magnitude ($M$). Among these structural relations, we have the Kormendy relation (KR; \cite{kor77}), the Faber-Jackson relation (FJR; \cite{fab76}) and the Fundamental Plane (FP; \cite{djo87}; \cite{dre87}), which are very useful tools for understanding the processes of formation and evolution of galaxies. In the recent past there have been many efforts directed to investigate the properties of ETGs using the structural relations. The large majority of these works have studied the coefficients of the structural relations at different wavelengths, environments, redshifts and luminosities and the results that they have found are very heterogeneous and discrepant. For example, several studies have found that the coefficients of the structural relations remain stable when considering different wavelengths (\cite{ben92}; \cite{BENDERETAL98}; \cite{ber03b}; \cite{ber03c}; \cite{lab05}; \cite{lab08}). However, other studies show that the structural relations depend on wavelength (\cite{jor96}; \cite{hud97}; \cite{pah98}; \cite{sco98}; \cite{jun08}). Several studies have shown that the structural relations and/or the structural parameters of galaxies are affected by the environment (\cite{ben92}; \cite{BENDERETAL98}; \cite{tru01}; \cite{tru02}; \cite{ber03b}; \cite{agu04}; \cite{gut04}; \cite{den05}; \cite{jor05}), although other studies have found the opposite conclusion (\cite{ros01}; \cite{tre01}; \cite{evs02}; \cite{gon03}; \cite{red04}; \cite{red05}; \cite{nig07}). When structural relations of galaxy samples are studied at different redshifts, several studies indicate that only the zero point of these relations depends on the redshift (\cite{bar98}; \cite{zie99}; \cite{lab03}; \cite{bar06}). However, other authors find that there is, not only a dependence of the zero point on redshift but that the slopes of the structural relations are steeper for higher redshift galaxies than for galaxies in the local Universe (\cite{tre05}; \cite{jor06}; \cite{fri09}). Finally, and regarding the luminosity, some authors find that dwarf and bright ellipticals follow structural relations with different coefficients (\cite{kor85}; \cite{ham87}; \cite{ben92}; \cite{cao93}; \cite{agu09}; \cite{des07}). Studies by Nigoche-Netro (2007) and Nigoche-Netro et al. (2008;2009;2010) find that the distribution of galaxies in the space of parameters that define the structural relations depends on the luminosity and that the coefficients of the structural relations depend on the width of the magnitude range ($\Delta M$) within which the galaxies are contained. They also find that when the width of the magnitude range diminishes, the differences in the slope of the structural relations (for intervals of the same width and different luminosity) become small and when it has approximately constant magnitude the differences are negligible. This effect is present in all samples of galaxies studied, independently of their degree of intrinsic difference, and has been denoted as geometrical effect.

On the other hand, the analysis of the intrinsic dispersion is less frequent in the astronomical literature than that of the values of the coefficients of the structural relations. Some papers have found that this dispersion depends both on luminosity (\cite{ben92}; \cite{jor96}; \cite{hyd09}) and the environment (\cite{ber03b}; \cite{den05}). The behaviour of the intrinsic dispersion of the FJR and the KR have not been studied thoroughly because it is considered that there is a third variable which causes most of the intrinsic dispersion in both these relations. The intrinsic dispersion of the FP has not been thoroughly studied because, it has been traditionally thought small ($\sim 0.1$ dex, \cite{kja93}, \cite{jor96}, \cite{kel97},  \cite{jor99}, \cite{bla02}, \cite{ber03c}, \cite{red05}, \cite{jor06}).  However, some studies show that the intrinsic dispersion values are far from being small ($\sim0.3$ dex, \cite{ben92}; \cite{lab03}; \cite{nig09}).
This, relatively high dispersion causes that the galaxy distribution in the space that defines the structural relations follows a surface whose thickness is determined by this dispersion.

Recent works (\cite{nig07b}; Nigoche-Netro et al. 2008; 2009; 2010) have shown that the intrinsic dispersion is also affected by the geometrical effect. However, Nigoche-Netro et al. (2010) have found that differences in the value of the intrinsic dispersion for different samples of galaxies do not disappear as the magnitude range diminishes.  They find that the exact value for the intrinsic dispersion is obtained when $\Delta M = 0$. The intrinsic dispersion for this extreme case would be defined as the standard deviation of the distribution of the points at constant magnitude. Hence, Nigoche-Netro et al. (2010) consider that an appropriate method for obtaining physical information for a sample of galaxies is to find its intrinsic dispersion at each magnitude value and then perform comparisons of this dispersion at different luminosities, wavelengths, redshifts, or environments. In light of this, in this paper we carry out a study of the behaviour of the intrinsic dispersion of the FJR as a function of luminosity, mass and redshift for a sample of ETGs selected from the SDSS-DR7 archive.

In \S~2 we present the galaxy sample used to study the intrinsic dispersion of the FJR, the calculation of the intrinsic dispersion and the analysis of the behaviour of the intrinsic dispersion as a function of the luminosity, mass and redshift. In \S~3 we present a discussion of the most important results of this paper. Finally in \S~4 we present our conclusions.

\section{The intrinsic dispersion of the structural relations for a sample of ETGs}

\subsection{The sample of ETGs}

We use a sample of ETGs from the Seventh Data Release of the SDSS (\cite{yor00}; \cite{aba09}) in $g$ and $r$ filters. This sample contains approximately 90 000 galaxies in each filter, distributed in a redshift interval $0.01 < z < 0.35$ and within a magnitude range $\Delta M\sim 7$ $mag$. The sample selection procedure was based on the Bernardi et al. (2003a) and Hyde \& Bernardi (2009) selection criteria (see \cite{nig10}). Hereinafter we will refer to it as the total SDSS sample. Given that the total sample spans a relatively ample redshift range, it is affected by the Malmquist bias. To avoid this bias, we use a volume-limited sample of approximately 17 000 ETGs with 0.04 $\leq\;z\;\leq$ 0.08 in the $g$ and $r$ filters. This subsample covers a magnitude range $<\Delta M>$ $\sim 4.5$ $mag$ ($-18.5 \ge M_{g} > -23.0$) in both filters and we refer to it as the homogeneous SDSS sample. This sample is complete approximately for $M_{g} \leq -20.0$.

\subsection{The structural relations of the ETGs and their intrinsic dispersion}

In recent papers (\cite{nig07b}; Nigoche-Netro et al. 2008; 2009; 2010) it has been demonstrated that due to the geometrical effect, it is risky to draw conclusions about the physical properties of galaxies by comparing the slopes of the structural relations for magnitude ranges of different widths or for magnitude ranges of the same width but of different luminosity, because, with the exception of the full magnitude interval, there is no ideal width at which comparisons should be made. So that using the slopes of the structural relations to find intrinsic differences among samples of galaxies is a delicate matter, and the results might be non conclusive. This procedure should be supplemented by an alternative corroborative method. Nigoche-Netro et al. (2010) demonstrate that the intrinsic dispersion values at constant magnitude is an appropriate method for obtaining physical information on a sample of galaxies. In the next sections we will use this method to try to find differences among the structural properties of galaxies belonging to different samples.
\subsection{The intrinsic dispersion of the FJR as function of the luminosity, mass and redshift}

\subsubsection{The intrinsic dispersion of the FJR as function of the luminosity}

Some papers from the literature have studied the intrinsic dispersion of the structural relations as a function of the luminosity (\cite{ben92}; \cite{jor96}, \cite{hyd09}; \cite{nig10}). Those works reveal that the intrinsic dispersion for bright galaxies is smaller than that for faint galaxies. However, these works have made comparisons of the intrinsic dispersion on wide magnitude ranges, so that, these results are affected by the geometrical effect. An appropriate analysis requires the calculation of the intrinsic dispersion at constant magnitude. In Figure 1 we show the behaviour of the intrinsic dispersion of the FJR ($\sigma_{log(\sigma_{0})}$) in very narrow magnitude ranges for the homogeneous and total samples from the SDSS in the $g$ filter. This figure shows the data for the homogeneous sample (Red Diamonds) where it is clearly seen that in the regime $M_{g} \lesssim -20$, where the sample is complete, the intrinsic dispersion value changes systematically as we consider brighter galaxies, and that the distribution of brighter galaxies presents a lower value for the intrinsic dispersion than the value presented by fainter galaxies. For the total SDSS sample (Black Dots), we observe the same behaviour.

We apply a non parametric statistical test (run test) in order to check a randomness hypothesis for our data sequence (see \cite{nig09}).  More precisely, it can be used to test the hypothesis that the data of the intrinsic dispersion of the FJR are mutually independent. With this test we find that there are reasons to affirm, with a 99\% level of confidence, that there is an underlying trend for the values of the intrinsic dispersion as a function of luminosity.

In order to characterise the behaviour of the intrinsic dispersion as a function of luminosity, we have fitted a straight line to those points in Figure 1 that correspond to the homogeneous sample at $M_{g} \le -20.0$. The resulting equation is:

\begin{equation}
\sigma_{log(\sigma_{0})}=(0.016 \pm 0.002)M_{g} + (0.434 \pm 0.029).
\end{equation}

The previous equation was obtained from a fit made with the {\it Bivariate Correlated Errors and Intrinsic Scatter bisector} ($BCES_{Bis}$)  (\cite{iso90}; \cite{akr96}) method. This method takes into consideration the errors in the variables, the error correlation, the data dispersion and both variables as dependent variables. This method is used for all the fits in this paper.

\subsubsection{The intrinsic dispersion of the FJR as function of the mass}

In this section, we analyse the behaviour of the intrinsic dispersion of the FJR as a function of the mass. We shall be using two different methods to calculate the mass of galaxies. The first method requires the galaxies' luminosity and colour indices and the following equation (see \cite{bel03}).

\begin{equation}
{\bf M_{g}} \sim L_{g} 10^{a_{g} + b_{g} (M_{g}-M_{r})},
\end{equation}

where ${\bf M_{g}}$ is the mass obtained from the luminosity in the $g$ filter ($L_{g}$), $M_{g}$ and $M_{r}$ are the magnitudes in the $g$ and $r$ filters, $a_{g}$ and $b_{g}$ are scale factors (see Table 7 from \cite{bel03}). From now on, the mass which we obtain from the luminosity shall be called the stellar mass.

The second method requires knowledge of the velocity dispersion, it also assumes that the galaxies are in virial equilibrium and
utilises the following equation:

\begin{equation}
{\bf M_{virial}} \sim  \frac{5 r_{e} \sigma_{0}^{2}}{G}.
\end{equation}

 where ${\bf M_{virial}}$ is the virial mass, $r_{e}$ is the effective radius, $\sigma_{0}$ is the central velocity dispersion and $G$ is the gravitational constant.

In Figure 2 we present a comparison of the mass obtained using both methods. In Figures 3 and 4 we see the relation between the mass and the velocity dispersion for the masses obtained with both methods. For a detailed discussion of Figs. 2, 3 and 4 see section 2.3.3.

In Figure 5, we show the behaviour of the virial mass as a function of redshift for galaxies contained in the SDSS total sample. Vertical lines represent the limits of the $0.04 \leq\;z\;\leq$ 0.08 redshift interval where the homogeneous sample of the SDSS is contained. We note that within these limits there is a deficiency of galaxies for log(${\bf M_{virial}}/{\bf M_{\odot}}) \lesssim 10.5$ (${\bf M_{\odot}}$ is the solar mass), so we may affirm that log(${\bf M_{virial}}/{\bf M_{\odot}}) = 10.5$ represents the approximate completeness limit of the homogeneous SDSS sample. On the other hand, the behaviour of the stellar mass as function of the redshift is similar to that of the virial mass, so that the approximate completeness limit for the homogeneous sample is, in this case, also log(${\bf M_{g}}/{\bf M_{\odot}}) = 10.5$

In studying the intrinsic dispersion as function of the mass, we require calculation of the intrinsic dispersion at constant mass in order to avoid the geometrical effect. In Figure 6 we show the behaviour of the intrinsic dispersion of the FJR in very narrow mass ranges (0.1-log(${\bf M_{virial}}/{\bf M_{\odot}}$) wide intervals) for the homogeneous and total samples from the SDSS. In this figure we see that the values of the FJR intrinsic dispersion depend on the virial mass, however, this mass was obtained from equation 3 which involves both the effective radius as well as the velocity dispersion, that is to say, there is a correlation between virial mass and the velocity dispersion which might affect the intrinsic dispersion estimate (see section 2.3.3 for more details). In order to avoid this possible bias, it is necessary to use the stellar mass. In Figure 7 we present the values of the FJR intrinsic dispersion as function of the stellar mass. This figure shows that the behaviour of the intrinsic dispersion as function of the stellar mass is similar to the behaviour of the intrinsic dispersion as function of the virial mass, in the sense that, the intrinsic dispersion value changes systematically as we consider more massive galaxies, and that more massive galaxies present a lower value for the intrinsic dispersion than the value for less massive galaxies. The run test confirms an underlying trend between the stellar mass and the intrinsic dispersion with a confidence level of approximately 99\%.

In order to characterise the behaviour of the intrinsic dispersion as a function of the virial mass, we have fitted the points of the homogeneous sample for log(${\bf M_{virial}}/{\bf M_{\odot}}) \ge 10.5$ in Figure 6 to a straight line whose equation is:

\begin{equation}
\sigma_{log(\sigma_{0})}=(-0.013 \pm 0.003)\; $log$({\bf M_{virial}}/{\bf M_{\odot}}) + (0.207 \pm 0.026).
\end{equation}

Similarly, in the case for stellar mass (Figure 7), we have fitted the homogeneous sample points for log(${\bf M_{g}}/{\bf M_{\odot}}) \ge 10.5$ to a straight line whose equation is:

\begin{equation}
\sigma_{log(\sigma_{0})}=(-0.041 \pm 0.005)\; $log$({\bf M_{g}}/{\bf M_{\odot}}) + (0.534 \pm 0.039).
\end{equation}

Equation 5 confirms that the correlation between virial mass and the velocity dispersion does not cause the behaviour of the intrinsic dispersion described by equation 4. Although this correlation could be behind the differences observed between the coefficients of both equations. In the following section we shall make an analysis of the possible origin of these differences.

\subsubsection{Differences between virial and stellar mass}

The difference between the coefficients in equations 4 and 5 may be due to various factors. One such factor is that virial and stellar mass might be intrinsically different (i.e. that the fit slope to both masses may be different from 1) and other factors would be associated with elements that would make the dispersion of virial mass as function of stellar mass to be relatively high (see Figure 2), and also, that the dispersion of the velocity dispersion with respect to virial mass, presents a different behaviour from that presented by the dispersion of the velocity dispersion with respect to stellar mass (see Figures 3 and 4). In what follows we shall analyse each one of these possibilities:

\begin{itemize}

\item 1) Differences due to the method to calculate the mass.
In accordance with equation 3, virial mass is composed both by the
stellar mass as well as by the dark mass. In our case, however, we see
that the contribution of the dark mass is not very important because
the slope of the virial mass vs. the stellar mass is approximately 1
(see Figure 2). So this does not explain the reported difference.

\item 2) Differences due to errors in the parameters involved in the mass estimates. In order to be able to explain the dispersion in the distribution of the stellar mass (see equation 2 and Figure 2) the errors in the magnitude (Filter g) should be of the order 1\%, however the mean errors in this magnitude are approximately of the order 0.1\%.
For the dispersion in the virial mass (see equation 3 and Figure 2), the error in the velocity dispersion  should be of the order 10\% , however, the mean error of this velocity dispersion is approximately 1\%. If the differences were due to the error in effective radius, this error should be of the order 20\%, however the mean effective radius error is of the order 4\%. An error combination in both variables might explain the differences. These errors should be at least of the order 5\% for the velocity dispersion and 10\% for the effective radius. From this we can say that the errors in the variables involved in the mass estimates are not responsible for the dispersion between the values of the virial and the stellar masses.

\item 3) Differences due to the correlation between virial mass, velocity dispersion and effective radius (see equation 3). If there would be a dispersion in the effective radius, not due to errors, this dispersion would affect the calculation of the virial mass but not that of the stellar mass, since this mass is independent from the effective radius value (see equation 2). This possibility may not be discarded a priori.

\item 4) Differences due to the lack of a parameter in the calculation of the virial or the stellar mass or in both. The dispersion in the masses might be due to a missing parameter, necessary in the calculation of either the virial or the stellar mass or in both. This possibility may not be discarded a priori.

\end{itemize}
Following what we have expressed above, cases 3 and 4 would play an important role in the behaviour of the intrinsic dispersion of these masses with respect to the velocity dispersion. Neither one of these cases may be discarded a priori, we only know that the final result of the action that one or the other may cause is different. Which value, then, is the correct one for the mass? or which one of equations 4 or 5 is the one that produces the appropriate results? The answer to this question is complicated and, at present, we do not have enough elements to elucidate as to what is the correct answer. As a consequence of this, we shall use both masses in order to investigate the universality of the FJR and that of the rest of the structural relations.

An interesting point that comes up when comparing virial and stellar mass (see 1 above) is that the slope of the fit is approximately equal to 1. This result is different from those of recent papers (\cite{cap06}; \cite{rob06}; \cite{koo10}) in which they find that the slope is up to 30\% larger, meaning that within the virial radius there is up to a 30\% of the total mass in the form of dark matter. What we find in this paper contradicts this result and shows that the amount of dark matter within the virial radius is negligible. We consider that, given its relevance, this result should be dealt with in a profound manner which we will do in a forthcoming paper, since this topic is beyond the scope of the present paper.

\subsubsection{The intrinsic dispersion of the FJR as function of the redshift}

An interesting property of the SDSS sample used in this work is that it contains galaxies in the redshift range $0.01 < z < 0.35$, so that, we can study the behaviour of the structural properties as function of redshift. In this case, and in order to avoid the geometrical effect, we can study the behaviour of the intrinsic dispersion of the FJR in two ways; the first one consists in considering samples of galaxies of approximately equal magnitude but with different redshift, and the second in considering samples of galaxies with approximately equal mass but with different redshift.

Figure 8 shows the behaviour of the intrinsic dispersion of the FJR as function of the redshift. Each symbol and colour represent a sample of galaxies of approximately constant magnitude. In some cases the errors and the narrow redshifts ranges where the samples are distributed make it difficult to observe any trend. However, for $M_{g} \sim -22.0$ and $M_{g} \sim -23.0$ we have data of relatively good quality so we are able to see a smooth trend (see Figure 9). The run test confirm an underlying trend with a confidence level of approximately 90\%.

In order to characterise the behaviour of the intrinsic dispersion as a function of the redshift, in Figure 9 we have fitted the points of the sample with $M_{g} \sim -22$ to a straight line whose equation is:

\begin{equation}
\sigma_{log(\sigma_{0})}=(-0.244 \pm 0.106)\; z + (0.109 \pm 0.025).
\end{equation}

Similarly the straight line equation obtained from the sample with $M_{g} \sim -23$ is:

\begin{equation}
\sigma_{log(\sigma_{0})}=(-0.248 \pm 0.113)\; z + (0.123 \pm 0.027).
\end{equation}

For the mass case, in Figure 10 we can see the behaviour of the intrinsic dispersion of the FJR as function of the redshift. In this graph each symbol and colour represent a sample of galaxies with approximately constant virial mass. In this case it also happens that the errors and the narrow redshifts ranges where some samples are distributed make it difficult to observe any trend. However, for the samples with log(${\bf M_{virial}}/{\bf M_{\odot}}) \sim 11.3$ and log(${\bf M_{virial}}/{\bf M_{\odot}}) \sim 11.8$, the data are of sufficiently good quality so that here we can see a clear trend (see Figure 11). The run test confirms an underlying trend with a confidence level of approximately 95\%. The intrinsic dispersion as function of the redshift has a similar behaviour when we use the stellar mass.

In order to characterise the behaviour of the intrinsic dispersion as a function of the redshift, in Figure 11 we have fitted the points of the sample with log(${\bf M_{virial}}/{\bf M_{\odot}}) \sim 11.3$ to a straight line whose equation is:

\begin{equation}
\sigma_{log(\sigma_{0})}=(-0.138 \pm 0.033)\; z + (0.076 \pm 0.009).
\end{equation}

Similarly, the straight line equation obtained from the sample with log(${\bf M_{virial}}/{\bf M_{\odot}}) \sim 11.8$ is:

\begin{equation}
\sigma_{log(\sigma_{0})}=(-0.150 \pm 0.020)\; z + (0.076 \pm 0.005).
\end{equation}

We must note that the dependence of the intrinsic dispersion of the FJR as a function of redshift has been found for the brightest and more massive galaxies, because in the faint and low mass end we do not have data or the ones that we do have are not of sufficiently good quality. On the other hand, the coefficients of equations 6 and 7 are compatible, within the errors, with the coefficients of equations 8 and 9. This behaviour may be understood by means of Figure 2 where we show that the difference between the virial and the stellar mass is smaller for larger values of the mass. Given that the objects we use for the study of the intrinsic dispersion as a function of the redshift are massive, the differences which might result should be relatively small.

The results presented in this section, allow us to affirm that, for the brighter and more massive galaxies, the intrinsic dispersion of the FJR changes systematically with distance. Those galaxies located further away, have a lower intrinsic dispersion of the FJR than those located closer by.

\section{Discussion}

The structural relations of ETGs, in particular the FP, have been studied extensively over the last 20-30 years. The physical explanation for the FP consider that the ETGs are in virial equilibrium and that they are homologous systems. The term homology means in this context the regular behaviour of both the mass-luminosity ratio and the structure along the entire range of ETGs luminosities. However, these assumptions are not sufficient to explain the observational results.
To try to find an explanation to the differences between theory and observation, which refer mainly to the tilt of the FP, several authors have invoked different mechanisms that affect the ETGs along their formation and evolution. One of the most important considerations in trying to explain the tilt is that the ETGs are non-homologous systems, however, up to the present the results of the different works (e.g. \cite{pah98}; \cite{sco98}; \cite{rob06}; \cite{jun08}) only explain partially the tilt, and there are no conclusive results.

A recent study (\cite{nig09}) has shown that the FP is not a simple plane in the space of parameters $\log(r_{e})$, $<\mu>_{e}$, $\log(\sigma_{0})$ and that the distribution of galaxies in this space depends on luminosity. In the case of the KR and FJR, Nigoche-Netro et al. (2008; 2010) find the same behaviour. Fraix-Burnet et al. (2010) have also found that the distribution of galaxies in the space of parameters is very complex, they established that the FP and other structural relations such as the FJR are formed by 7 different groups of galaxies. These groups define different separate regions on the graphical planes or space where the structural relations are plotted. They also find that each group is truly `homologous', where `homology', for them, means similarity due to having the same class of progenitor. Each group defines its own structural relation which is more loosely defined for less-diversified groups. The `diversity' means, in this context, the number and nature of transformation events that affect the galaxies along their life time such as: collapse, accretion, interaction and merging. So that the term less-diversified (more-diversified) means that galaxies have suffered fewer (more) transformation events along their life times.

It is important not to confuse the terms `diversity' and `diversification'. `Diversification', following Fraix-Burnet et al. (2006), refers to the number of different classes of objects that are present in a sample. An illustrative example of the difference between these two concepts is the following: Four identical galaxies when mixed in pairs would produce two different galaxies. Different among themselves and different from the original ones. If these two galaxies would mix again, the mixture would produce a new class of galaxy. In this example, along the entire process, 4 different classes of objects have been produced, but the final result is that there is only one class of galaxy and this last object is the more diversified, because is the one that has suffered the larger number of transformation events.

Fraix-Burnet et al. (2010) conclude that the FP could be an historical and not a physical correlation, so that, the correlation among $\log(r_{e})$, $<\mu>_{e}$ and $\log(\sigma_{0})$ might not be a tilted virial plane due to dissipation or to a particular behaviour of ${\bf M}/L$, but rather a parametric correlation between the evolution of these parameters. In other words, the FP would be the result of several transforming events such as collapse, accretion, interaction and merging, that is to say, the FP would be the result of the historical sequence of events that affected, one way or another, the physical structure of the galaxies in question.

In previous sections we have characterised the variation of the distribution of the ETGs in the plane of the parameters $M$ and $\log(\sigma_{0})$ using the intrinsic dispersion.  We have found that the intrinsic dispersion of the FJR depends on the luminosity and mass of the galaxies and that the distribution of brighter and more massive galaxies has a lower intrinsic dispersion than that for the fainter and less massive galaxies. This result is in agreement with the work of Fraix-Burnet et al. (2010), where, in their Figure 7 we can see that the region of the brighter galaxies on the FJR plane is formed by the more diversified groups -only one or two groups populate this region-, while the region of the fainter galaxies is formed by the less diversified groups -four or five groups populate this region-, which occupy a more ample region. Put differently, the distribution of galaxies inside the groups and the distribution of the groups on the FJR plane might cause that the distribution of brighter galaxies, which appear to be those more diversified, have a lower intrinsic dispersion than that of the distribution for the fainter galaxies, which seem to be the less diversified ones. So, we may conclude that the intrinsic dispersion of the FJR might depend on several transforming events such as collapse, accretion, interaction and merging, in other words, just as for the FP, the intrinsic dispersion of the FJR might be a consequence of the historical series of events which affected the physical structure of the galaxies.

We have also found that, for bright and massive galaxies in the redshift range $0.01 < z < 0.25$, the intrinsic dispersion of the FJR depends on the redshift and that the distribution of the farther galaxies has a lower intrinsic dispersion than that for the nearer galaxies. Fraix-Burnet et al. (2010) have not found a relation between their groups and redshift, however, the redshift range that they use is very narrow ($0.007 < z < 0.053$), so that the dependence of groups on redshift might have gone unnoticed.

The results which we have mentioned above are preliminary, given that the sample used in this paper and that of Fraix-Burnet et al. (2010) are different. The former sample corresponds to ETGs with different environments and redshifts, while the latter corresponds to ETGs in nearby clusters. Prior to taking these results as conclusive, it is necessary to perform an analysis of the diversification of the galaxies for samples in different environments and with different redshifts. It is also important to note that the behaviour of the intrinsic dispersion, which we have found, might be due to the involvement of other phenomena. Phenomena that the Fraix-Burnet et al. (2010) analysis has not contemplated, such as considering that the dispersion might have a component inherent to the objects from their origin, and not only due to the transformation events that they have undergone along their lifetimes.

\section{Conclusions}

Analysing the intrinsic dispersion of the FJR for a sample of approximately 90 000 ETGs from the literature we find the following:

\begin{itemize}

\item The intrinsic dispersion of FJR depends on the luminosity, mass and redshift.

\item The distribution of brighter and more massive galaxies has a lower intrinsic dispersion than the distribution for the fainter and less massive galaxies.

\item In the redshift range $0.01 < z < 0.25$, for luminous and massive galaxies, the distribution of the farther galaxies has a lower intrinsic dispersion than that for the nearer galaxies.

\end{itemize}

A recent work (\cite{fra10}) has shown that the structural relations, among which we have the FJR, are formed by 7 different groups of galaxies. These groups define separate regions on the structural relations. Each group defines its own structural relation which is more loosely defined for less-diversified groups. The `diversity' means in this context the number and the nature of transformation events that affect the galaxies along their life times. When we compare our results with those of Fraix-Burnet et al. (2010) we find that the distribution of galaxies inside the groups and the distribution of the groups on the FJR plane might cause that the brighter (more-diversified) galaxies have lower intrinsic dispersion than the fainter (less-diversified) galaxies. So we may conclude that the intrinsic dispersion of the FJR might depend on the number and the nature of transformation events that affect the galaxies such as collapse, accretion, interaction and merging, that is to say, the intrinsic dispersion might depend on the histories of galaxies.

As mentioned above (Section 3), before taking the former results as conclusive, further diversification studies should be carried out in ETGs samples similar to the one presented in this paper. It is also important to consider that the intrinsic dispersion might depend on other factors which have not been considered in the analysis by Fraix-Burnet et al. (2010) such as the fact that part of this dispersion might be due to original inherent properties of the galaxies and not only to the transformation events which they have suffered along their life times.

Finally, as seen in section 2.3, a comparison of virial and stellar mass returns a value of the slope approximately equal to 1, this value differs from recent results in the literature in which the slope could be as large as $\sim 1.3$. This would mean that inside the virial radius 30\% of the mass could be in the form of dark matter. Our results indicate that inside this radius the amount of dark matter is negligible. We shall discuss this point in a forthcoming paper.

\begin{acknowledgements}

We would like to dedicate this humble work to the memory of Mrs. Eutiquia Netro Castillo, an extraordinary woman.

We would like to thank Consejo Nacional de Ciencia y Tecnolog\'{\i}a (M\'{e}xico) for a postdoctoral fellowship number 117993, Instituto de Astrof\'{\i}sica de Andaluc\'{\i}a (IAA, Espa\~na), Instituto de Astrof\'{\i}sica de Canarias (IAC, Espa\~na) and Instituto de Astronom\'{\i}a (UNAM, M\'exico) for all the facilities provided for the realisation of this project. We acknowledge financial support from projects: CSD2006 00070 "1st Science with GTC", of the CONSOLIDER 2010 Programme; and
AYA2010-21887-C04, AYA2007-67965-C03, "Estallidos", PNAYA of the Spanish MICINN. We would also like to thank Jana Benda for her help with this paper.

\end{acknowledgements}

\clearpage

\begin{figure*}[!htb]
\centering
\includegraphics[angle=0,width=12cm]{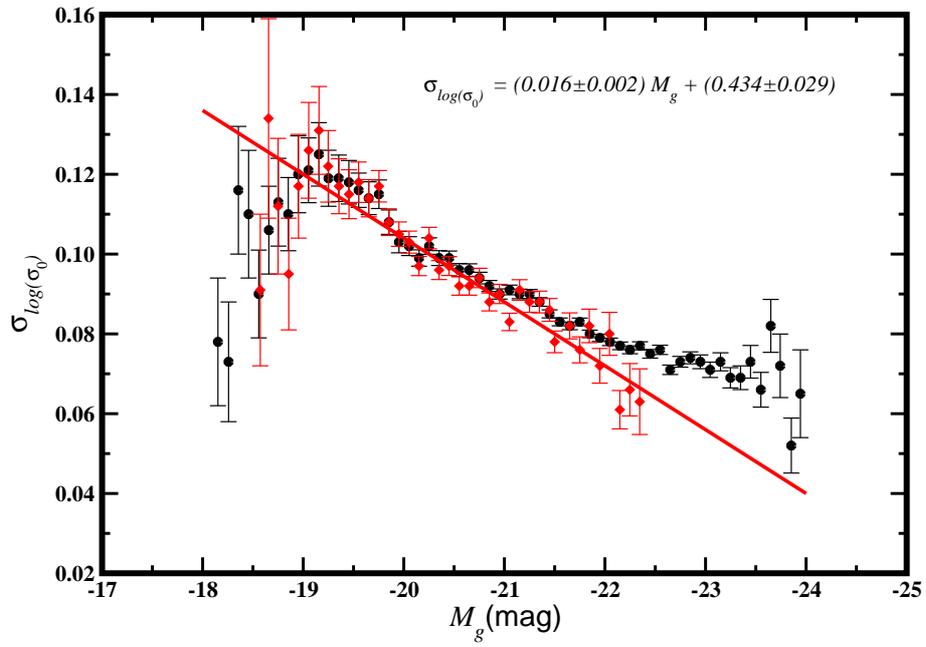}
\caption{Variation of the FJR intrinsic dispersion ($\sigma_{log(\sigma_{0})}$) in 0.1-$mag$ wide intervals. Circles represent the total SDSS sample ($g$ filter) and diamonds the homogeneous SDSS sample ($g$ filter). The straight line represents a fit ($BCES_{Bis}$) to the points in the homogeneous sample brighter than $M_{g} \sim -20$.}
\end{figure*}

\clearpage

\begin{figure*}[!htb]

\psfrag{O}{\Large $\odot$}

\centering
\includegraphics[angle=0,width=12cm]{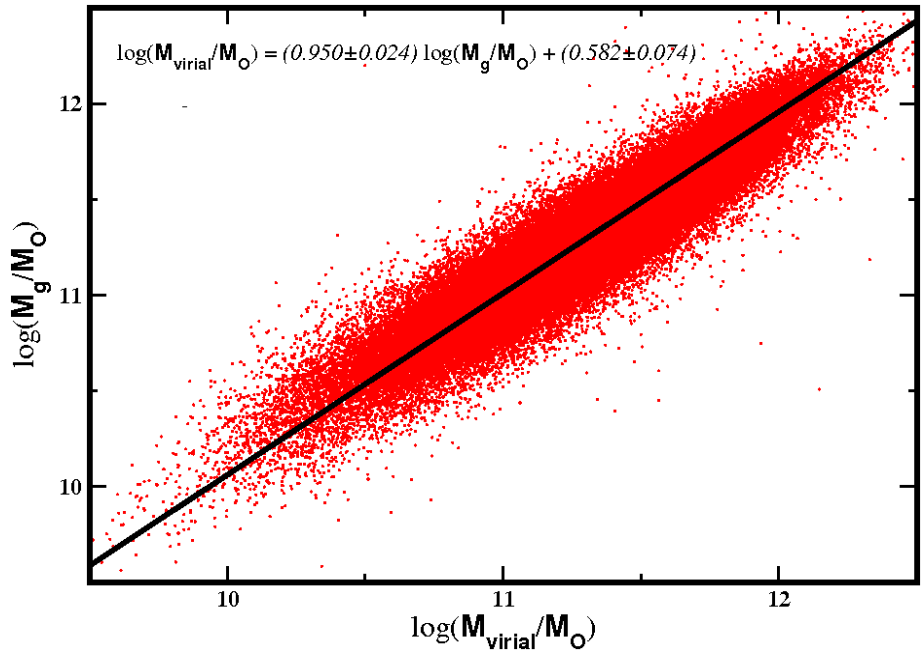}
\caption{Comparison of the mass obtained from the velocity dispersion and the mass obtained from the luminosity. The black line represents a fit ($BCES_{Bis}$) to the points in the total sample.}
\end{figure*}

\clearpage

\begin{figure*}[!htb]

\psfrag{O}{\Large $\odot$}

\centering
\includegraphics[angle=0,width=12cm]{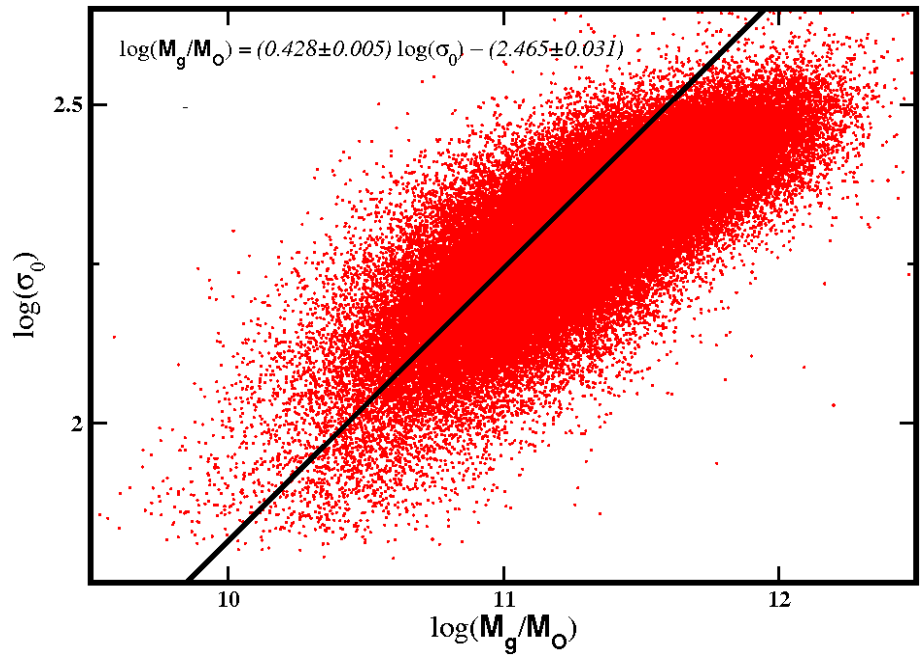}
\caption{Faber-Jackson relation using the mass obtained from the luminosity. The black line represents a fit ($BCES_{Bis}$) to the points in the total sample.}
\end{figure*}

\clearpage

\begin{figure*}[!htb]

\psfrag{O}{\Large $\odot$}

\centering
\includegraphics[angle=0,width=12cm]{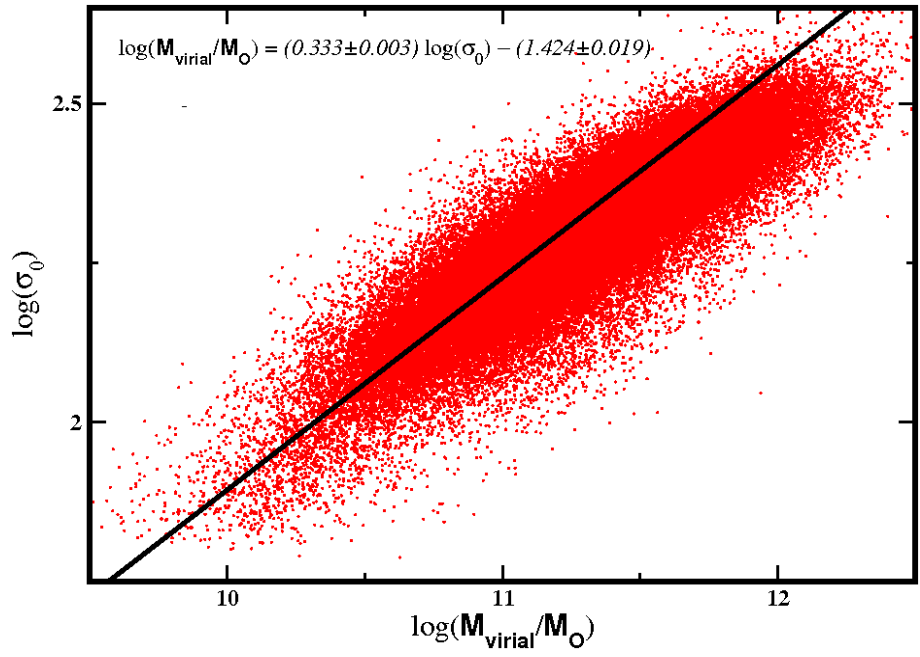}
\caption{Faber-Jackson relation using the mass obtained from the velocity dispersion. The black line represents a fit ($BCES_{Bis}$) to the points in the total sample.}
\end{figure*}

\clearpage

\begin{figure*}[!htb]

\psfrag{O}{\Large $\odot$}

\centering
\includegraphics[angle=0,width=12cm]{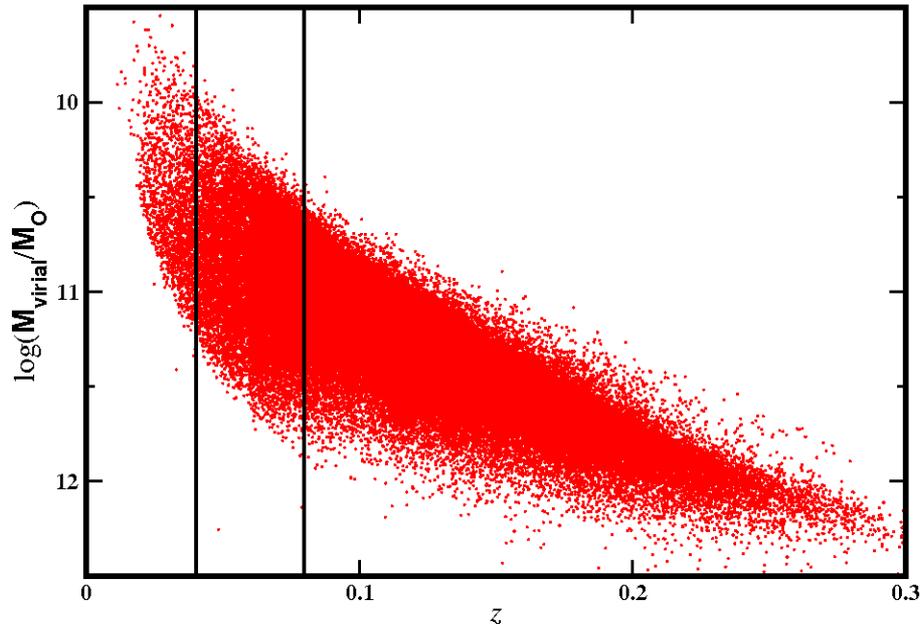}
\caption{Distribution of the mass as function of redshift for the total SDSS sample. The vertical lines correspond to the 0.04 $\leq\;z\;\leq$ 0.08 redshift interval within which the homogeneous sample is contained. In this interval, we note that for log(${\bf M_{virial}}/{\bf M_{\odot}}) \lesssim 10.5$ there exists a deficiency of galaxies, so log(${\bf M_{virial}}/{\bf M_{\odot}}) = 10.5$ could be considered to be the completeness limit of the homogeneous SDSS sample.}
\end{figure*}

\clearpage

\begin{figure*}[!htb]

\psfrag{O}{\Large $\odot$}

\centering
\includegraphics[angle=0,width=12cm]{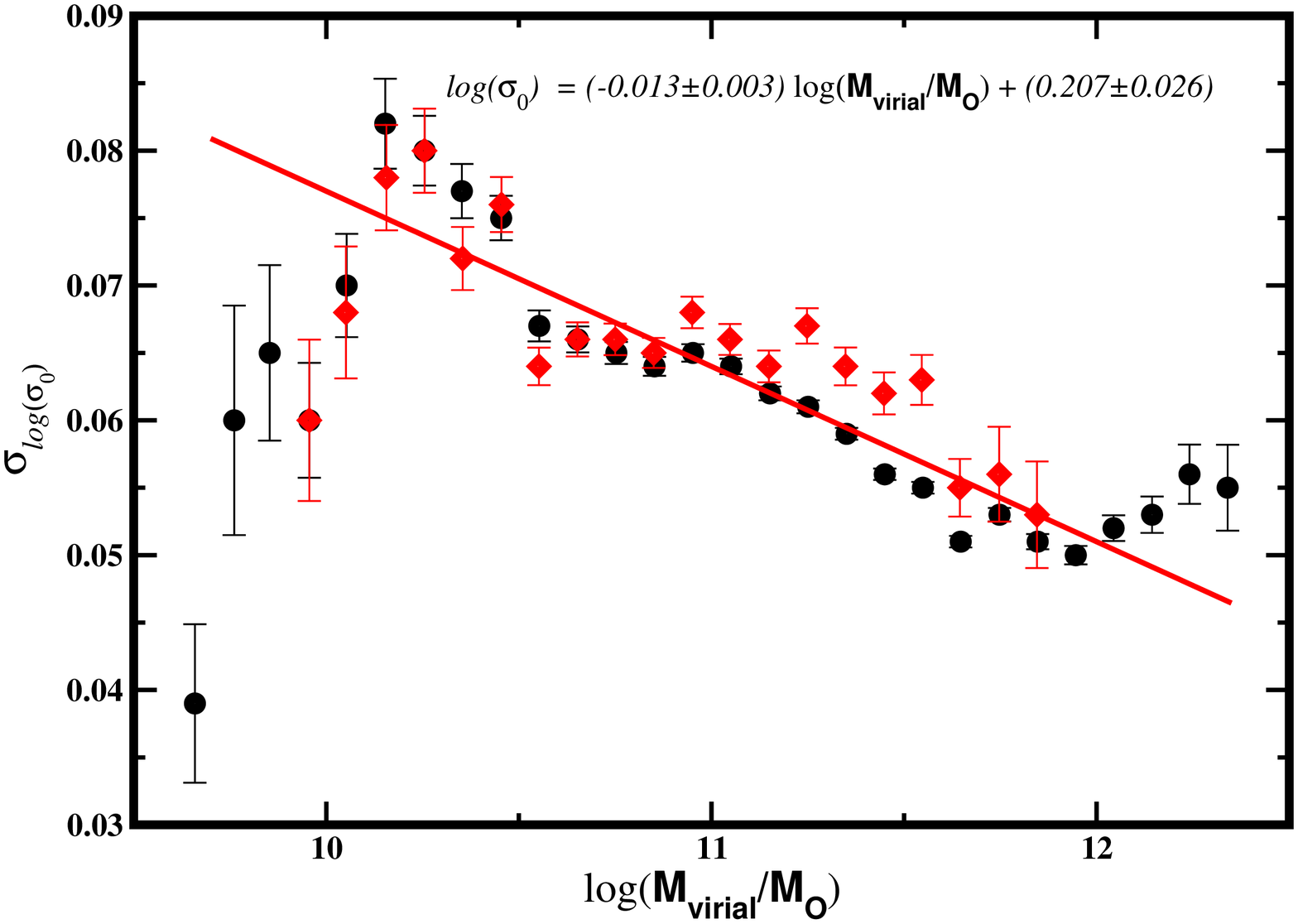}
\caption{Variation of the FJR intrinsic dispersion ($\sigma_{log(\sigma_{0})}$) in 0.1-log(${\bf M_{virial}}/{\bf M_{\odot}}$) wide intervals. Circles represent the total SDSS sample and diamonds the homogeneous SDSS sample. The straight line represents a fit ($BCES_{Bis}$) to the points in the homogeneous sample for log(${\bf M_{virial}}/{\bf M_{\odot}}) \ge 10.5$. }
\end{figure*}

\clearpage

\begin{figure*}[!htb]

\psfrag{O}{\Large $\odot$}

\centering
\includegraphics[angle=0,width=12cm]{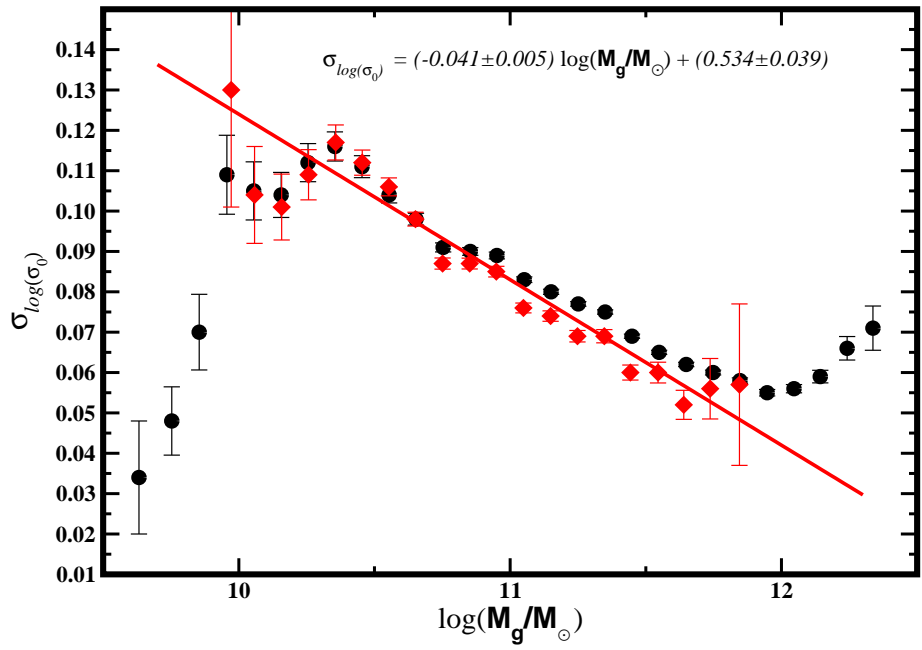}
\caption{Variation of the FJR intrinsic dispersion ($\sigma_{log(\sigma_{0})}$) in 0.1-log(${\bf M_{g}}/{\bf M_{\odot}}$) wide intervals. Circles represent the total SDSS sample and diamonds the homogeneous SDSS sample. The straight line represents a fit ($BCES_{Bis}$) to the points in the homogeneous sample for log(${\bf M_{g}}/{\bf M_{\odot}}) \ge 10.5$. ${\bf M_{g}}$ represents the stellar mass.}
\end{figure*}

\clearpage

\begin{figure*}[!htb]
\centering
\includegraphics[angle=0,width=12cm]{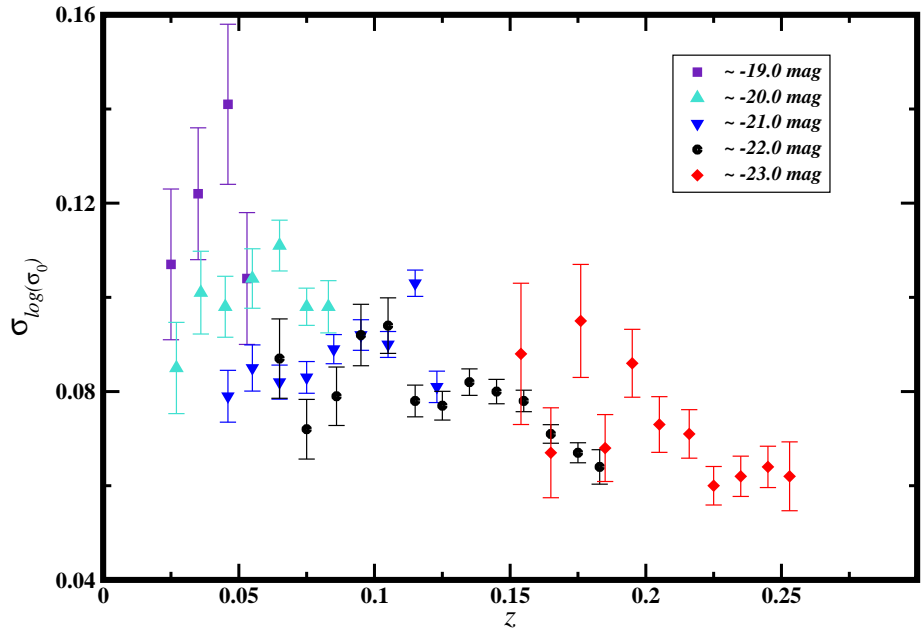}
\caption{Intrinsic dispersion of the FJR ($\sigma_{log(\sigma_{0})}$) as function of redshift. Each symbol and colour represent an approximately constant magnitude ($g$ filter).}
\end{figure*}

\clearpage

\begin{figure*}[!htb]
\centering
\includegraphics[angle=0,width=12cm]{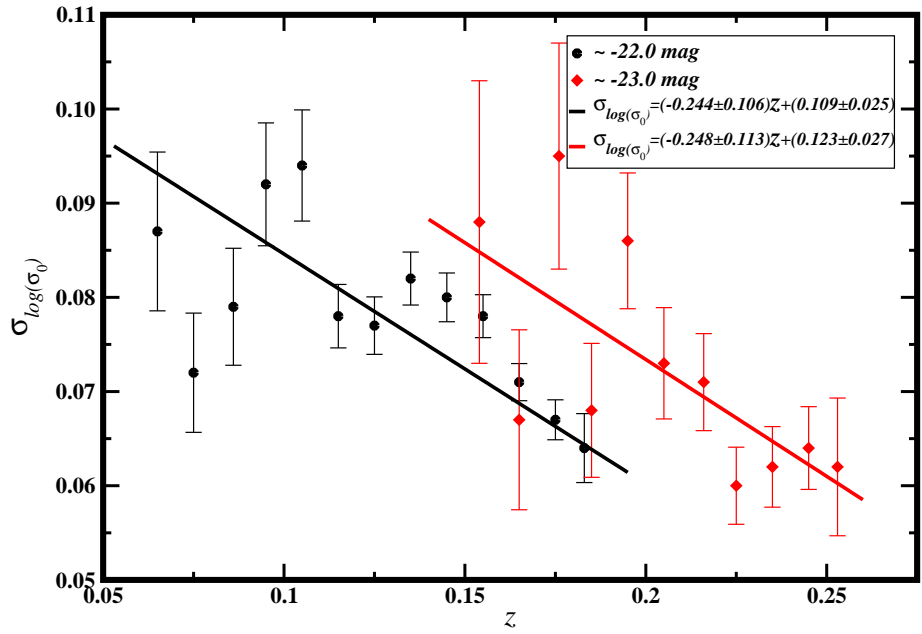}
\caption{Intrinsic dispersion of the FJR ($\sigma_{log(\sigma_{0})}$) as function of redshift. Each symbol and colour represent an approximately constant magnitude ($g$ filter). The straight lines represent fits ($BCES_{Bis}$) to all the points in each sample.}

\end{figure*}

\clearpage

\begin{figure*}[!htb]
\centering
\includegraphics[angle=0,width=12cm]{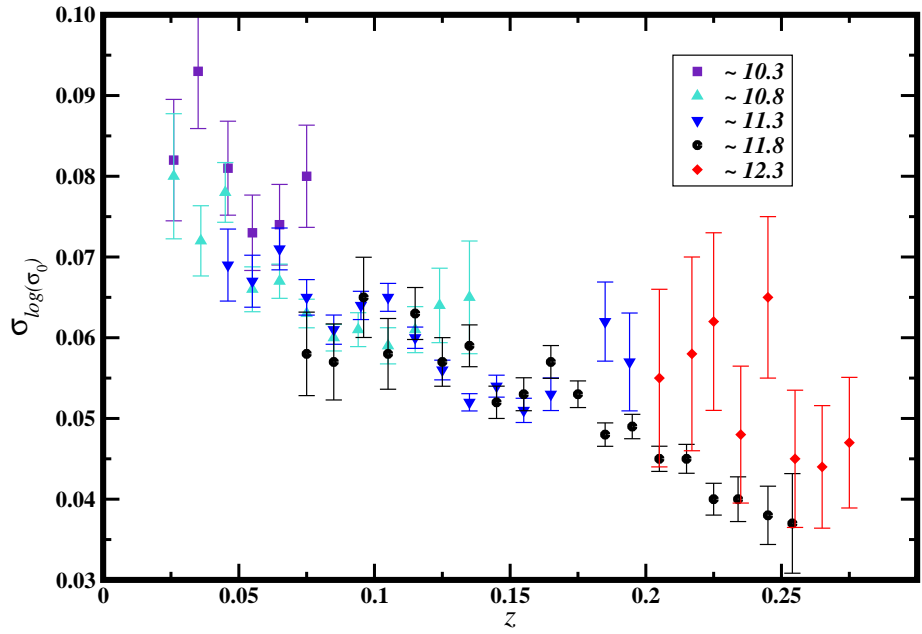}
\caption{Intrinsic dispersion of the FJR ($\sigma_{log(\sigma_{0})}$) as function of redshift. Each symbol and colour represent an approximately constant virial mass.}
\end{figure*}

\clearpage

\begin{figure*}[!htb]
\centering
\includegraphics[angle=0,width=12cm]{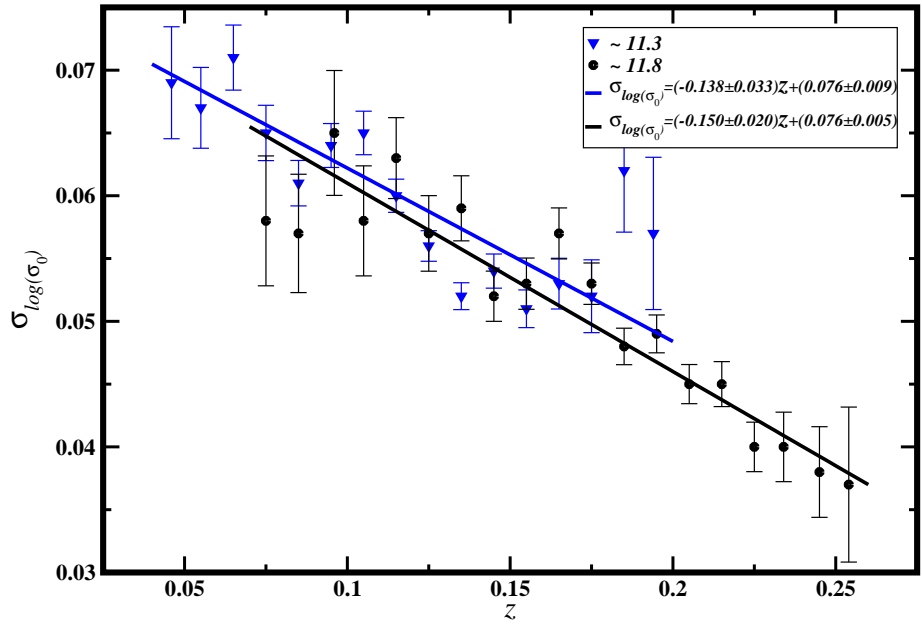}
\caption{Intrinsic dispersion of the FJR ($\sigma_{log(\sigma_{0})}$) as function of redshift. Each symbol and colour represent an approximately constant virial mass. The straight lines represent fits ($BCES_{Bis}$) to all the points in each sample. }
\end{figure*}



\end{document}